\documentclass[showpacs,aps,nofootinbib,floatfix,amsmath,amssymb,prd]{revtex4}
\usepackage{graphicx}
\begin{document}


\title{Bilepton effects on the $WWV^*$ vertex  in the 331 model with right-handed neutrinos
via a $SU_L(2)\times U_Y(1)$ covariant quantization scheme}
\author{F. Ram\'\i rez-Zavaleta}
\author{G. Tavares-Velasco}
\author{J. J. Toscano} \affiliation{Facultad de
Ciencias F\'{\i}sico Matem\'aticas, Benem\'erita Universidad
Aut\'onoma de Puebla, Apartado Postal 1152, Puebla, Pue., M\'exico}

\date{\today}

\begin{abstract}
In a recent paper \cite{Montano:2005gs}, we investigated the effects
of the massive charged gauge bosons (bileptons) predicted by the
minimal 331 model on the off-shell vertex $WWV^*$ ($V=\gamma$, $Z$)
using a $SU_L(2)\times U_Y(1)$ covariant gauge-fixing term for the
bileptons. We proceed along the same lines and calculate the effects
of the gauge bosons predicted by the 331 model with right-handed
neutrinos. It is found that the bilepton effects on the $WWV^*$
vertex are of the same order of magnitude than those arising from
the SM and several of its extensions, provided that the bilepton
mass is of the order of a few hundred of GeVs. For heavier
bileptons, their effects on the $WWV^*$ vertex are negligible.  The
behavior of the form factors at high energies is also discussed as
it is a reflect of the gauge invariance and gauge independence of
the $WWV^*$ Green function obtained via our quantization method.
\end{abstract}


\maketitle

\section{Introduction}

Radiative corrections to the  $WWV$ ($V=\gamma, Z$) vertex have long
received considerable attention. Apart from its sensitivity to new
physics effects, this vertex has theoretical interest by its own as
it may serve as a probe of the gauge sector of the standard model
(SM). In this context, the one-loop contributions to the $WW\gamma$
vertex, which defines the static electromagnetic properties of the
$W$ boson have been calculated in the SM
\cite{Bardeen:1972vi,Argyres:1992vv} and several of is extensions
such as two-Higgs doublet models (THDMs) \cite{Couture:1987eu},
supersymmetric models \cite{Bilchak:1986bt}, 331 models
\cite{Tavares-Velasco:2001vb,Garcia_Luna:2004}, etc. Similar
attention has been paid to the study of the $WWZ$ vertex. Despite
the attention has focused mainly on the static properties of the $W$
boson, a more comprehensive study of the $W$ boson properties
requires the analysis of the off-shell $WWV$ vertex, particularly
when the neutral $V$ boson is off-shell and the $W$ bosons are
on-shell as the resulting vertex could be tested with high precision
at the CERN large hadron collider (LHC) or the planned future linear
colliders. It is well known that although the on-shell $WWV$ vertex
renders a gauge independent amplitude by itself, gauge independence
is lost once any external particle goes off-shell, thereby requiring
the use of a nonconventional calculation scheme. The difficulties to
obtain a gauge independent Green function for the $WWV^*$ vertex
have long been known. In particular, a careless calculation of the
radiative corrections to the $WWV^*$ vertex via a conventional
gauge-fixing procedure irremediably leads to an ill-behaved gauge
dependent Green function. For instance, one-loop corrections to the
$WWV^*$ vertex were first calculated in the SM by the authors of
Ref. \cite{Argyres:1992vv} using the Feynman-'t Hooft gauge via a
conventional gauge-fixing scheme. The result was an infrared
divergent Green function that violates unitarity. Aware of the fact
that an off-shell Green function is generally gauge dependent,
authors of Ref. \cite{Papavassiliou:1993ex} invoked a diagrammatic
method known as the pinch technique (PT) \cite{PT} to obtain a gauge
independent $WWV^*$ Green function satisfying the requirements of
gauge invariance and infrared finiteness. The PT exploits the fact
that although the off-shell Green functions are gauge dependent, the
$S$-matrix elements to which they contribute are gauge independent.
In this way, order by order in perturbation theory, one can
construct a gauge independent Green function for an off-shell vertex
by combining its contribution with  all other Feynman diagrams that
also contribute to a particular physical process, thereby getting
rid of any $\xi$ dependent terms. In this respect, while the $WWV^*$
vertex does not represent a physical process by its own, it can
contribute to a gauge invariant physical process such as $e^+e^-\to
W^+W^-$ scattering. By this method, a gauge invariant and gauge
independent $WWV^*$ Green function can be constructed. The PT was
also used in Ref. \cite{Arhrib:1995dm} to calculate the one-loop
contributions to the $WWV^*$ vertex from the unconstrained minimal
supersymmetric standard model (MSSM).

The PT is thus a diagrammatic method that allows one to remove any
unphysical gauge dependent term from an off-shell $n$-point function
at the level of the Feynman graphs contributing to a certain
physical process, thereby yielding a well-behaved gauge-independent
and gauge-invariant Green function. The PT may turn itself, however,
into a somewhat cumbersome task. There is also the alternative of
removing any gauge dependent term at the level of the generating
functional. In this respect, although the background field method
(BFM) \cite{BFM} renders a gauge invariant quantum action, no
mechanism has been found yet that allows one to obtain both
gauge-invariant and gauge-independent Green functions from the
generating functional. Eventually, a connection between the PT and
the BFM has been found \cite{BFMPTLink}: it turns out that the
gauge-invariant and gauge-independent Green function obtained via
the PT is exactly reproduced if it is calculated through the BFM
using the Feynman rules given in the Feynman-'t Hooft gauge
($\xi=1$). Although such a correspondence, which so far remains a
puzzle, was first established at the one-loop level
\cite{Papavassiliou:1994yi}, it has been shown that it persists at
all orders of perturbation theory \cite{Binosi:2002ft}. Therefore,
instead of using the PT, one can use the BFM Feynman-'t Hooft gauge
(BFMFG) to calculate gauge-independent off-shell amplitudes. Along
these lines, in Ref. \cite{Montano:2005gs} we calculated the
one-loop contributions to the $WWV^*$ vertex from the gauge sector
of the so called minimal 331 model
\cite{Pisano:1991ee,Frampton:1992wt}. In order to construct a
well-behaved gauge independent Green function, rather than using the
PT, we invoked an alternative method inspired in the BFM and BRST
symmetry \cite{Becchi:1974md}, which is well suited to study the
sensitivity of the $WWV^*$ vertex to the virtual effects of the new
gauge bosons predicted by 331 models. We will follow a similar
approach here to calculate the one-loop contributions from the 331
model with right-handed neutrinos \cite{Foot:1994ym}. Although we
will shortly describe the calculation scheme below, for more details
we refer the reader to Ref. \cite{Montano:2005gs}.

331 models have been the source of great interest lately.  Recent
studies within the framework of this class of theories have focused
on neutrino physics \cite{Ponce:2006au}, $Z'$ physics
\cite{Barreto:2007mt}, Higgs boson physics \cite{Liu:2007yk},
bilepton physics
\cite{Montano:2005gs,Garcia_Luna:2004,Tavares-Velasco:2004be,Honorato:2006nr},
supersymmetric extensions \cite{Dong:2007qc}, dark matter
\cite{Long:2006ab}, CP violation \cite{Promberger:2007py}, and
theoretical aspects \cite{Dias:2006ns}. Models of this kind are
based on the ${SU_L(3)}\times {U_X(1)}$ gauge symmetry
\cite{Pisano:1991ee,Frampton:1992wt} and are unique in the sense
that they require that all the 3 fermion families be summed up in
order to cancel anomalies, in contrast with other models in which
anomaly cancelation is achieved family by family. As a consequence,
331 models require that the number of fermion families be a multiple
of 3, the quark color number. This may suggest a solution to the
family replication problem. Apart from this feature, 331 models are
interesting as they predict a pair of massive gauge bosons arranged
in a doublet of the electroweak group. While the minimal 331 model
predicts a pair of singly-charged $Y^\pm$ and a pair of
doubly-charged $Y^{\pm\pm}$ gauge bosons, the model with
right-handed neutrinos predicts a pair of neutral no self-conjugate
gauge bosons $Y^0$ instead of the doubly-charged ones. These new
gauge bosons are called bileptons since they carry two units of
lepton number, thereby being responsible for lepton-number violating
interactions \cite{Foot:1994ym}. The neutral bilepton has been
deemed  a promising candidate in accelerator experiments since it
may be the source of neutrino oscillations \cite{Long:1999yv}. The
reason why the effects of the bileptons on the $WWV^*$ vertex are
worth studying is because their couplings to the SM gauge bosons are
rather similar to the SM gauge boson couplings. Current bounds put
the bilepton masses in the range of a few hundred GeVs
\cite{Ky:2000dj}, which means that the bileptons may show up via
their virtual effects in low-energy observables. This is an
important reason to investigate the effect of these particles on the
$WWV^*$ vertex. Furthermore, due to the spontaneous symmetry
breaking (SSB) hierarchy, the splitting between the charged and
neutral bilepton masses is bounded by $m_{W}$, so the bilepton
masses might be almost degenerate since they are expected to be
heavier than  $m_{W}$. We will see below that this is particularly
suited for our calculation method.

The rest of the paper is organized as follows. In Sec. II we present
a brief introduction to the 331 model with right-handed neutrinos. A
survey of the quantization method is presented in Sec. III along
with a detailed discussion on the gauge fixing procedure for the
bileptons, whereas Sec. IV is devoted to the analytical results and
the analysis. Finally, the conclusions are presented in Sec. V.

\section{The 331 model with right-handed neutrinos}

The 331 model with right-handed neutrinos was first introduced in
Ref. \cite{Foot:1994ym}. In a previous paper, we worked in the
context of this model and calculated the static electromagnetic
properties of the $W$ boson \cite{Garcia_Luna:2004}. Also, the
one-loop contributions to the static electromagnetic properties of
the neutral no self-conjugate $Y^0$ boson were calculated in Ref.
\cite{Tavares-Velasco:2004be}. It was shown that the main
contributions to the $WW\gamma$ vertex arise from the gauge sector,
i.e. from the bileptons, as the fermion sector does not contribute
and the Higgs sector gives a negligibly contribution quite similar
to that arisen in a THDM \cite{Garcia_Luna:2004}. Therefore,
although for completeness we will present a short description of the
main features of the model, the following discussion will be mainly
focused on the gauge sector as it is the one which is expected to
give the dominant contributions to the $WWV^*$ vertex. Furthermore,
we will see that our calculation scheme is suited to analyze the
effects of the bileptons on the $WWV^*$ Green function.

In the fermion sector of the 331 model with right-handed neutrinos,
leptons of the same generation are arranged in left-handed triplets
and right-handed singlets, the same being true for each quark
generation. Apart from the introduction of right-handed neutrinos,
there are three new quarks $D_1$, $D_2$, and $T$, which have the
following electric charge $Q_{D_i}=-1/3\, e$ and $Q_T=2/3\, e$. In
order to cancel the $SU_L(3)$ anomaly, two quark families transform
as $SU_L(3)$ antitriplets and the remaining one as a triplet. At the
first stage of SSB, when the $SU_L(3)\times U_N(1)$ group is
spontaneously broken into $SU_L(2)\times U_Y(1)$, the new quarks get
their masses and emerge as singlets of the electroweak group.
Consequently, they cannot interact with the $W$ boson at the tree
level. It follows that there is no contribution from the new quarks
to the $WWV^*$ vertex at the one-loop level.

As far as the scalar sector is concerned, several Higgs sectors have
been proposed in the literature to achieve the SSB in 331 models
\cite{Tonasse:1996cx}. As for the 331 model with right-handed
neutrinos, it requires the simplest Higgs sector of this class of
theories \cite{Foot:1994ym}, i.e. only three triplets of $SU_L(3)$
are required to reproduce the SM physics at the Fermi scale:

\begin{eqnarray}
\chi=\left( \begin{array}{ccc} \Phi_Y \\
\chi^{'0}
\end{array}\right)\sim (1,3,-1/3),\ \
\rho=\left( \begin{array}{ccc} \Phi_1 \\
\rho^{'+}
\end{array}\right)\sim (1,3,2/3),\ \
\eta=\left( \begin{array}{ccc} \Phi_2 \\
\eta^{'0}
\end{array}\right)\sim (1,3,-1/3),
\end{eqnarray}
where $\Phi^\dag_Y=(G^{0*}_Y,G^+_Y)$,
$\Phi^\dag_1=(\rho^-,\rho^{0*})$ and
$\Phi^\dag_2=(\eta^{0*},\eta^+)$ are $SU_L(2)\times U_Y(1)$ doublets
with hypercharge -1, 1 and -1 respectively. The vacuum expectation
values (VEV) are $<\chi>^T=(0,0,w/\sqrt{2})$,
$<\rho>^T=(0,u/\sqrt{2},0)$, and $<\eta>^T=(v/\sqrt{2},0,0)$. The
triplet $\chi$ breaks the $SU_L(3)\times U_N(1)$ group into
$SU_L(2)\times U_Y(1)$ at the $w$ scale, whereas $\rho$ and $\eta$
are meant to break $SU_L(2)\times U_Y(1)$ into $U_e(1)$.

In the gauge sector, the model predicts the existence of five new
gauge bosons: two singly charged $Y^\pm$, two neutral no
self-conjugate $Y^0$, and a neutral self-conjugate $Z'$. All these
new gauge bosons and the new quarks as well acquire their masses at
the $w$ scale.  At this stage of SSB, the $Z'$ boson emerges as a
singlet of the electroweak group and so no interaction between the
$Z'$ boson and the $W$ boson arises. However, in the following stage
of SSB, at the Fermi scale, the $Z'$ boson couples with the $W$
boson via the $Z-Z'$ mixing angle. As a consequence, the $Z'$
contribution to the $WWV^*$ vertex will be proportional to the
square of the $Z-Z'$ mixing angle, which is highly suppressed
according to the most recent experimental bounds
\cite{Liu:1993fw,Ng:1992st}. We will thus ignore this contribution
in this work. In sharp contrast, the bileptons arise as a doublet of
the electroweak group at the $w$ scale, which means that they couple
to the SM gauge bosons in a rather peculiar way: due to the fact
that the $SU_L(2)$ group is completely embedded in $SU_L(3)$, the
bileptons couplings to the SM gauge bosons have similar strength and
Lorentz structure than those of the SM gauge boson couplings
themselves. These new interactions, which arise solely from the
Yang-Mills sector, do not involve any mixing angle and are entirely
determined by the $SU_L(2)$ coupling constant and the weak angle
$\theta_W$.

In the first stage of SSB, the bilepton masses are degenerate:
$m_{Y^0}=m_{Y^\pm}=m_Y=gw/2$.  However, when $SU_L(2)\times U_Y(1)$
is broken down to $U_e(1)$, the bilepton masses degeneration is also
broken. Once the Higgs kinetic-energy sector is diagonalized, there
emerge the following mass-eigenstate fields:

\begin{eqnarray}
&&Y^0_\mu=\frac{1}{\sqrt{2}}(A^4_\mu-iA^5_\mu), \\
&&Y^-_\mu=\frac{1}{\sqrt{2}}(A^6_\mu-iA^7_\mu),\\
&&W^+_\mu=\frac{1}{\sqrt{2}}(A^1_\mu-iA^2_\mu),
\end{eqnarray}
with masses $m^2_{Y^0}=g^2(w^2+u^2)/4$,
$m^2_{Y^\pm}=g^2(w^2+v^2)/4$, and $m^2_W=g^2(u^2+v^2)/4$. The
remaining three gauge fields $A^3_\mu$, $A^8_\mu$,and $N_\mu$ define
the self-conjugate mass eigenstates. From the above expressions, it
is straightforward to obtain the following upper bound on the
bilepton masses splitting \cite{Foot:1994ym}:
\begin{equation}
\label{splitting} \left|m^2_{Y^0}-m^2_{Y^\pm}\right|\leq m^2_W.
\end{equation}

Therefore, the bilepton masses become nearly degenerate when they
are much larger than $m_W$. While in the minimal 331 model an upper
bound on the bilepton masses of the order of 1 TeV can be derived
from the theoretical constraint $\sin^2 \theta_W\leq 1/4$
\cite{Frampton:1992wt,Ng:1992st,Dias:2004dc}, which is obtained from
matching the gauge couplings constants at the $SU_L(3)\times U_X(1)$
breaking scale, in the version with right-handed neutrinos the
corresponding bound is highly relaxed as the theoretical constraint
is $\sin^2\theta_W\leq 3/4$ \cite{Foot:1994ym}.

The Yang-Mills sector induces all the couplings we need to compute
the bilepton contribution to the $WWV^*$ vertex, so we will devote
particular attention to it. The respective Lagrangian can be written
as

\begin{equation}
{\cal L}_{YM}=-\frac{1}{4}F^a_{\mu \nu}F^{\mu
\nu}_a-\frac{1}{4}N_{\mu \nu}N^{\mu \nu},
\end{equation}
where $F^a_{\mu \nu}=\partial_\mu A^a_\nu-\partial_\nu
A^a_\mu+f^{abc}A^b_\mu A^c_\nu$ and $N_{\mu \nu}=\partial_\mu
N_\nu-\partial_\nu N_\mu$, being $f^{abc}$ the structure constants
associated with $SU_L(3)$. When the mass eigenstate fields are
introduced, the above Lagrangian can be split into three
$SU_L(2)\times U_Y(1)$-invariant terms \cite{Foot:1994ym}:
\begin{equation}
\label{L_YM} {\cal L}_{YM}={\cal L}^{SM}_{YM}+{\cal
L}^{SM-NP}_{YM}+{\cal L}^{NP}_{YM}.
\end{equation}
\noindent While ${\cal L}^{SM}_{YM}$ stands for the usual SM
Yang-Mills Lagrangian, ${\cal L}^{SM-NP}_{YM}$ comprises the
interactions between the new gauge bosons and the SM ones. The
latter Lagrangian is the only one relevant for the present
discussion and we will get back to it later. As for the last term,
${\cal L}^{NP}_{YM}$, it induces the couplings of the $Z'$ boson to
the bileptons. At the one-loop level there are no contributions to
the $WWV^*$ vertex induced by this Lagrangian.

In order to calculate the one-loop correction to the $WWV^*$ vertex,
we will introduce a $SU_L(2)\times U_Y(1)$-covariant gauge-fixing
procedure for the bileptons. This will be discussed in the following
section.

\section{A $SU_L(2)\times U_Y(1)$ covariant gauge-fixing procedure for the bileptons}
\subsection{Overview of the quantization method}

We now turn to present an overview of the quantization method used
to obtain a gauge invariant and gauge independent Green function for
the $WWV^*$ vertex in the 331 model with right-handed neutrinos.
This method, which is inspired in the BFM and BRST formalism, was
comprehensively discussed in our previous work \cite{Montano:2005gs}
and we refrain from presenting a detailed discussion here.

The BFM \cite{BFM} is a powerful tool that allows one to construct a
gauge invariant quantum action out of which gauge invariant Green
functions can be obtained that are free of pathologies and satisfy
simple Ward identities. In a conventional quantization formalism,
all the fields appearing in the action are quantized. In the BFM,
the gauge fields are decomposed into a quantum field and a classical
(background) field. While the quantum fields are integrated out, the
background fields are treated as sources. As a result, the quantum
fields can only appear as internal lines in loop diagrams, whereas
the background fields appear as external lines.  In principle, it is
necessary to gauge-fix both the quantum and the classical fields in
order to define $S$-matrix elements, but it is enough to gauge-fix
the former to obtain off-shell Green functions. In this respect,
while gauge invariance with respect to the quantum fields is broken
when they are gauge-fixed, this process leaves unaltered the gauge
invariance with respect to the classical fields. The resulting Green
functions will be gauge invariant but they will still maintain the
dependence on the gauge parameter that characterizes the
gauge-fixing scheme used for the quantum fields. However, we can
exploit the connection between the BFM and the PT to obtain a gauge
invariant and gauge independent Green function. We just need to use
the BFM Feynman rules given in the Feynman-'t Hooft gauge. We will
see below that our quantization method incorporates some features of
the BFM.

Although gauge invariance with respect to the full theory is broken
when a conventional quantization scheme is applied, one can still
preserve the gauge invariance under a subgroup of the theory. This
approach is well suited when the virtual effects of the gauge fields
associated with the subgroup are expected to be considerably small.
In this context, following the philosophy of the effective
Lagrangian approach, where a $SU_L(2)\times U_Y(1)$ effective
Lagrangian is constructed out of the light (SM) fields to assess the
effects of the heavy fields on a low-energy observable, it would be
convenient to analyze the virtual effects of the bileptons on the
$WWV^*$ Green function in a $SU_L(2)\times U_Y(1)$-covariant manner.
A quantization scheme for the bileptons would only be required since
the SM gauge fields would only appear as external lines. Thus, a
$SU_L(2)\times U_Y(1)$-invariant effective Lagrangian can be
constructed by introducing a $SU_L(2)\times U_Y(1)$-covariant
gauge-fixing procedure  for the bileptons, which then can be
integrated out. The gauge-fixing procedure must involve the
$SU_L(2)\times U_Y(1)$-covariant derivative given in the
representation in which the heavy fields transform under this group,
which is the reason why such gauges are known as nonlinear or
covariant. This class of gauges have proved to be very useful for
the calculation of radiative corrections in the SM and beyond
\cite{NLGSM}.

Since we want to introduce a nonlinear gauge fixing procedure for
the bileptons, we need to be careful as the difficulties to
implement the Faddeev-Popov method (FPM) \cite{Faddeev:1967fc} in
such a case have long been known. More specifically, it is known
that renormalizability becomes ruined when the FPM is applied to a
nonlinear gauge.  We can invoke instead the BRST formalism
\cite{Becchi:1974md} to construct the most general renormalizable
nonlinear gauge-fixing term \cite{Toscano}. We will thus introduce a
$SU_L(2)\times U_Y(1)$-covariant gauge-fixing term for the
bileptons, which will lead to  an invariant quantum action out of
which a gauge invariant $WWV^*$ Green function will be obtained.
Invoking the connection between the PT and the BFM, the use of the
Feynman-'t Hooft Feynman rules will render a gauge independent Green
function.

Finally, we would like to point out that both our quantization
method and the BFM are meant to construct gauge-invariant quantum
actions. The main difference resides in the fact that while the BFM
allows one to analyze any new physics effects regardless the energy
scale, ours is only appropriate to study heavy physics effects on
low-energy (SM) Green functions. This stems from the fact that while
in the BFM gauge-invariance would be preserved with respect to the
gauge group of a complete theory, in our quantization method there
is only invariance under a subgroup of such a theory. In the case of
the present paper, our quantization method preserves gauge
invariance under the $SU_L(2)\times U_Y(1)$ group rather than
$SU_L(3)\times U_N(1)$. While the calculation will be greatly
simplified because of electroweak invariance, the price to be paid
is that our result will only be approximate as it will only account
for the bilepton effects on the $WWV^*$ vertex at the $w$ scale,
when the bilepton masses are still degenerate. We will see below
that this is indeed a good approximation.

\subsection{Gauge-fixed Lagrangian and Feynman rules}

The procedure to gauge-fix the Yang-Mills Lagrangian of the 331
model with right-handed neutrinos is similar to that described in
the case of the minimal 331 model. So, we refrain from presenting a
detailed discussion here and refer the reader to Ref.
\cite{Montano:2005gs}.

After introducing the most general action for a Yang-Mills system
consistent with BRST symmetry and renormalization theory
\cite{Batalin:1984jr}, we integrate out the auxiliary fields to
obtain an action defined by the following gauge-fixed $SU_L(2)\times
U_Y(1)$-invariant Lagrangian
\begin{equation}
{\cal L}_{BRST}={\cal L}_{YM}+{\cal L}_{GF}+{\cal L}_{FP},
\end{equation}
where  ${\cal L}_{GF}$ and ${\cal L}_{FP}$ are the gauge-fixing term
and the ghost sector Lagrangian, respectively. As for the Yang-Mills
Lagrangian ${\cal L}_{YM}$, which is given in Eq. (\ref{L_YM}), for
our calculation we only need the ${\cal L}^{SM-NP}_{YM}$ term as it
is the only one that induces the interactions between the bileptons
and the SM gauge bosons. It can be expressed as
\begin{eqnarray}
\label{L_SM-NP} {\cal L}^{SM-NP}_{YM}&=&-\frac{1}{2}\left(D_\mu
Y_\nu-D_\nu Y_\mu\right)^\dag \left(D^\mu Y^\nu-D^\nu Y^\mu\right)
-iY^\dag_\mu\left( g{\bf F}^{\mu \nu} +g'{\bf B}^{\mu
\nu}\right)Y_\nu\nonumber \\
&-&\frac{ig}{2}\frac{\sqrt{3-4s^2_W}}{c_W}
Z^\prime_{\mu}\Big(Y^\dag_\nu \left(D^\mu Y^\nu-D^\nu
Y^\mu\right)-\left(D^\mu Y^\nu-D^\nu Y^\mu\right)^\dag Y_\nu \Big),
\end{eqnarray}
where $Y^\dag_\mu=(Y^{0*}_\mu, Y^+_\mu)$ is a doublet of the
electroweak group with hypercharge $-1$, and
$D_\mu=\partial_\mu-ig{\bf A}_\mu +ig'{\bf B}_\mu$ is the
electroweak covariant derivative. We have also introduced the
definitions ${\bf F}_{\mu \nu}=\sigma^iF^i_{\mu \nu}/2$, ${\bf
A}_\mu=\sigma^iA^i_\mu/2$, and ${\bf B}_\mu=YB_\mu/2$, with
$\sigma^i$ the Pauli matrices.

As for the gauge-fixing term ${\cal L}_{GF}$ and the  ghost sector
Lagrangian ${\cal L}_{FP}$, they can be written as
\begin{equation}
{\cal L}_{GF}=-\frac{1}{2\xi}f^{\bar{a}}f^{\bar{a}},
\end{equation}
and
\begin{equation}
{\cal L}_{FP}=-\bar{C}^{\bar{a}}(\delta
f^{\bar{a}})-\frac{2}{\xi}f^{\bar{a}bc}f^{\bar{a}}\bar{C}^bC^c-\frac{1}{2}f^{\bar{a}bc}
f^{cde}\bar{C}^{\bar{a}}\bar{C}^bC^dC^e,
\end{equation}
where $f^{\bar{a}}$ are the gauge-fixing functions and
$\bar{C}^{\bar{a}}$ stands for the antighost fields. In addition,
$C^a$ are the ghost fields associated with the $A^a_\mu$ fields,
$f^{abc}$ are the $SU_L(3)$ structure constants, and $\xi$ is the
gauge parameter.

According to Ref. \cite{Montano:2005gs}, the most general
$SU_L(2)\times U_Y(1)$-covariant gauge-fixing functions
$f^{\bar{a}}$ consistent with renormalization theory is given by:
\begin{equation}
f^{\bar{a}}=(\delta^{\bar{a}b}\partial_\mu-gf^{\bar{a}bi}A^i_\mu)A^{\mu
b}-\frac{\xi g}{\sqrt{3}}f^{\bar{a}b8}\chi^\dag \lambda^b \chi, \ \
\ \bar{a}=4,5,6,7; \ \ \ i=1,2,3,8.
\end{equation}

We can now insert this expression into the gauge-fixed Lagrangian.
Apart from analyzing the dynamics induced by each term of the
gauge-fixed Lagrangian, we would like to put special emphasis on the
covariance under $SU_L(2)\times U_Y(1)$.

The covariant structure of the gauge-fixing term becomes manifest
when the mass eigenstates $f^{0}_Y=\frac{1}{\sqrt{2}}(f^4-if^5)$ and
$f^{-}_Y=\frac{1}{\sqrt{2}}(f^6-if^7)$ are introduced in an
$SU_L(2)\times U_Y(1)$ doublet
\begin{equation}
f_{Y}=\left(\begin{array}{ccc} f^{0}_Y \\
f^-_Y
\end{array}\right)=\Big(D_\mu-\frac{ig\sqrt{3-4s^2_W}}{2c_W}Z'_\mu\Big)Y^\mu-
\frac{ig\xi}{\sqrt{2}}\chi^{'0*}\Phi_Y.
\end{equation}
We will now decompose the gauge-fixing Lagrangian into three terms
in order to analyze the dynamics it induces:
\begin{equation}
{\cal L}_{GF}={\cal L}_{GF1}+{\cal L}_{GF2}+\dots,
\end{equation}
where each term is separately $SU_L(2)\times U_Y(1)$ invariant:
\begin{eqnarray}
{\cal L}_{GF1}&=&-\xi^{-1}(D_\mu Y^\mu)^\dag (D_\nu
Y^\nu)-\frac{\xi g^2}{2}(\chi^{'0*}\chi^{'0})(\Phi^\dag_Y\Phi_Y), \\
{\cal L}_{GF2}&=&\frac{ig}{\sqrt{2}}\left(\chi^{'0*}(D_\mu
Y^\mu)^\dag \Phi_Y-\chi^{'0}\Phi^\dag_Y(D_\mu Y^\mu)\right),
\end{eqnarray}
whereas the third term, denoted by $\dots$, involves the $Z'$ boson
and is  irrelevant here.

\begin{figure}[!htb]
\centering
\includegraphics[height=4.in,width=5.9in]{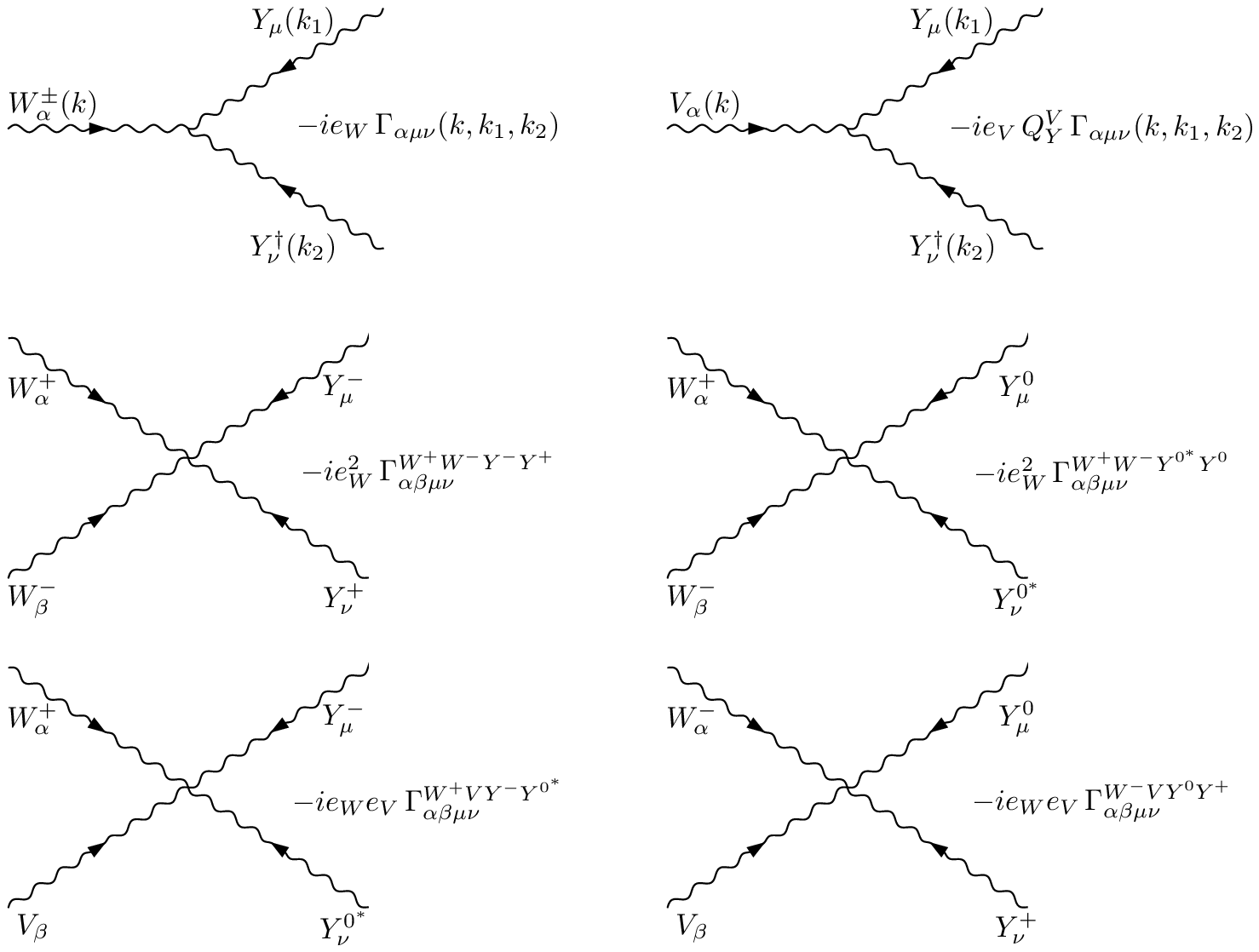}
\caption{\label{FIG1} Feynman rules for the trilinear and quartic
vertices involving the bileptons and SM gauge fields in the
$SU_L(2)\times U_Y(1)$-covariant $R_\xi$-gauge. $e_V=e$,
$Q^V_{Y^-}=-1$, and $Q^V_{Y^{0}}=0$ for $V=\gamma$, whereas
$e_V=g/(2c_W)$, $Q^V_{Y^-}=2s^2_W-1$, and $Q^V_{Y^{0}}=1$ for $V=Z$.
In addition, $e_W=g/\sqrt{2}$. See Eqs. (\ref{Gamn})-(\ref{GWVYY})
for the Lorentz structures.}
\end{figure}

We note that the first term of ${\cal L}_{GF1}$ defines the bilepton
propagators and gives new contributions to the trilinear and quartic
couplings involving the  bileptons and the SM gauge bosons, which
originally arise from the ${\cal L}^{SM-NP}_{YM}$ Lagrangian. After
some algebra, we are lead to the Feynman rules for these modified
couplings, which are shown in Fig. \ref{FIG1}.  The Lorentz
structure associated with the trilinear couplings is given by
\begin{equation}
\label{Gamn} \Gamma_{\alpha \mu
\nu}(k,k_1,k_2)=\left(k_2-k_1\right)_\alpha g_{\mu
\nu}+\left(k-k_2-\xi^{-1} k_1\right)_\mu g_{\alpha
\nu}-\left(k-k_1-\xi^{-1} k_2\right)_\nu g_{\alpha \mu},
\end{equation}
whereas those of the quartic couplings are
\begin{equation}
\label{GWWYY} \Gamma^{WWYY}_{\alpha \beta \mu \nu}=g_{\alpha
\beta}g_{\mu \nu}-2g_{\alpha \nu}g_{\beta
\mu}+\left(1+\xi^{-1}\right)g_{\alpha \mu}g_{\beta \nu},
\end{equation}
and
\begin{equation}
\label{GWVYY} \Gamma^{WVYY}_{\alpha \beta \mu
\nu}=(Q^V_{Y^-}+Q^V_{Y^{0}})g_{\alpha \beta}g_{\mu
\nu}+\left[\left(1+\xi^{-1}\right)Q^V_{Y^0}-
2Q^V_{Y^{-}}\right]g_{\alpha\mu}g_{\beta\nu}
+\left[\left(1+\xi^{-1}\right)Q^V_{Y^-}-
2Q^V_{Y^{0}}\right]g_{\alpha\nu}g_{\beta\mu}.
\end{equation}

The reason why the trilinear couplings $WYY$ and $VYY$ have the same
Lorentz structure is a consequence of $SU_L(2)\times U_Y(1)$
invariance. This is also reflected in the fact that a simple Ward
identity is satisfied by these vertices:
\begin{equation}
k^\alpha \Gamma_{\alpha \mu \nu}(k,k_1,k_2)=\Pi^{Y^\dag Y^\dag}_{\mu
\nu}(k_2)-\Pi^{YY}_{\mu \nu}(k_1),
\end{equation}
where $\Pi^{YY}_{\mu \nu}(k)$ is the two-point vertex function
\begin{equation}
\Pi^{YY}_{\mu \nu}(k)=(-k+m^2_Y)g_{\mu
\nu}-\left(\xi^{-1}-1\right)k_\mu k_\nu.
\end{equation}

Note that the unphysical masses for the pseudo-Goldstone bosons
associated with the bileptons  are determined by the scalar part of
${\cal L}_{GF1}$, which also modifies various unphysical couplings
arising from the Higgs potential. As for ${\cal L}_{GF2}$, it allows
one to remove various unphysical vertices that arise from the $\chi$
kinetic-energy sector ${\cal L}_{K\chi}$, which can be written as:
\begin{equation}
{\cal L}_{K\chi} ={\cal L}_{K\chi 1}+{\cal L}_{K\chi 2}+\dots,
\end{equation}
where each term is $SU_L(2)\times U_Y(1)$-invariant by itself:
\begin{equation}
{\cal L}_{K\chi 1}=(D_\mu \Phi_Y)^\dag (D^\mu \Phi_Y)+\partial_\mu
\chi^{'0*}\partial^\mu
\chi^{'0}+\frac{g^2}{2}\left(\chi^{'0*}\chi^{'0}Y^\dag_\mu Y^\mu
+(\Phi^\dag_YY_\mu)(Y^{\mu \dag}\Phi_Y)\right),
\end{equation}
\begin{equation}
{\cal L}_{K\chi2}=ie_W\left(\chi^{'0*}Y^\dag_\mu (D^\mu
\Phi_Y)+\Phi^\dag_YY_\mu
\partial^\mu \chi^{'0}-{\rm H.c.}\right)
\end{equation}
Once again, the terms denoted by $\dots$  are irrelevant for our
calculation as they involve the $Z'$ boson.

\begin{figure}
\centering
\includegraphics[height=3.5in,width=5.5in]{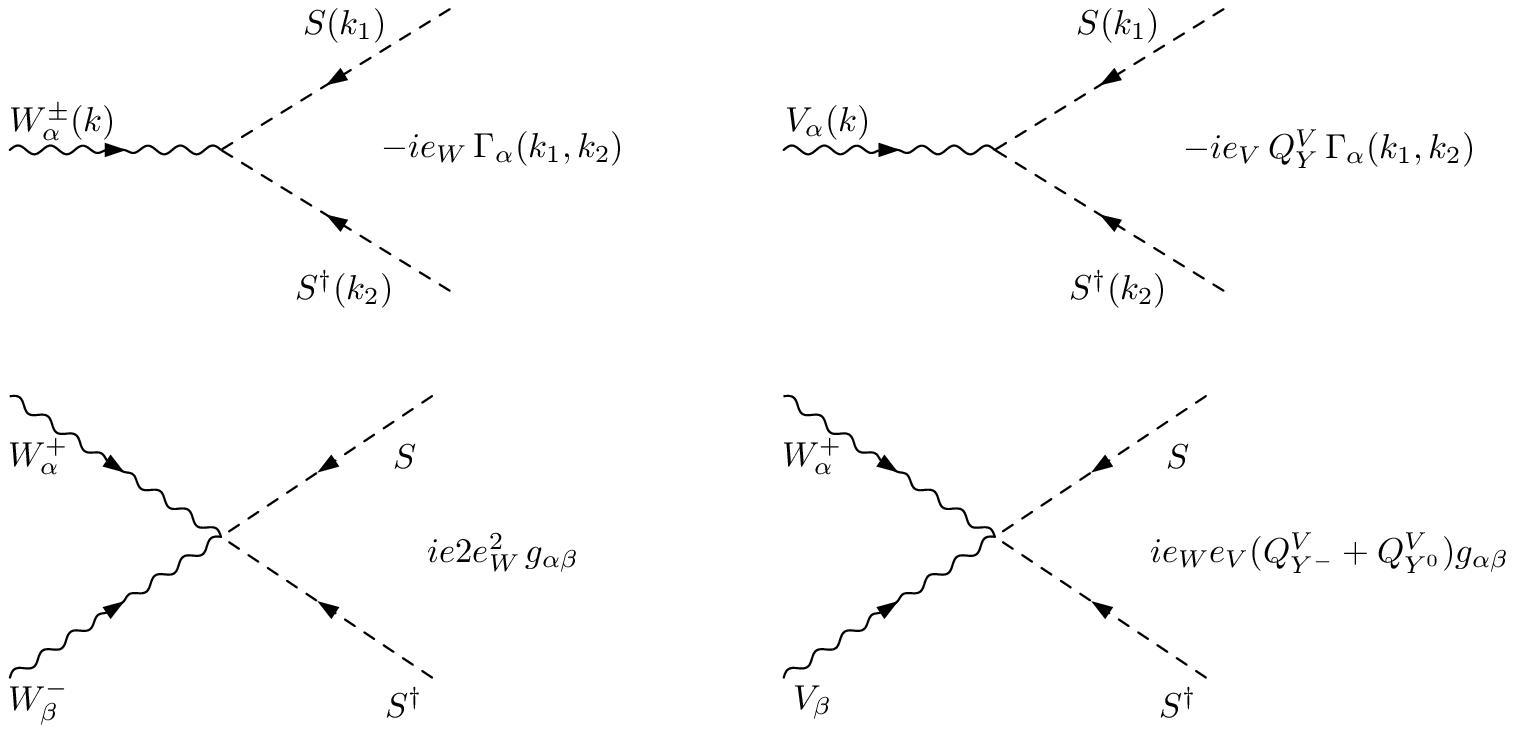}
\caption{\label{SFR} Feynman rules for the trilinear and quartic
vertices involving SM gauge fields and scalar unphysical particles
(pseudo-Goldstone bosons and ghosts) in the $SU_L(2)\times
U_Y(1)$-covariant $R_\xi$-gauge. In this gauge, the $W$ and $V$
couplings to pseudo-Goldstone bosons and ghosts coincide.}
\end{figure}

From these expressions it is clear that ${\cal L}_{K\chi1}$ induces
the couplings of the pseudo-Goldstone bosons to the SM gauge bosons.
The respective Feynman rules for these coupling can be extracted
straightforwardly and are presented in Fig. \ref{SFR}. As for ${\cal
L}_{K\chi2}$, it induces the bilinear terms $Y^{0,0*}_\mu
G^{0*,0}_Y$ and $Y^{\pm }_\mu G^{\mp }_Y$, together with the
unphysical trilinear couplings $Y^{0,0*}_\mu W^{-,+} G^{+,-}_Y$,
$Y^{+,-}_\mu W^{-,+} G^{0,0*}_Y$, and some quartic vertices
irrelevant for our calculation. It turns out that all these
couplings exactly cancel when $ {\cal L}_{K\chi2}$ and ${\cal
L}_{GF2}$ are combined:
\begin{equation}
 {\cal L}_{K\chi2} +{\cal
L}_{GF2}=ie_W\left(\chi^{'0*}\partial_\mu (Y^{\mu
\dag}\Phi_Y)+\Phi^\dag_YY_\mu \partial^\mu\chi^{'0}-{\rm
H.c.}\right)+\dots,
\end{equation}
where $\dots$ stands for irrelevant surface terms. This is the
reason why this gauge-fixing procedure will simplify considerably
our calculation as we can get rid of several Feynman diagrams
involving the couplings $Y^{0,0*}_\mu W^{-,+} G^{+,-}_Y$ and
$Y^{+,-}_\mu W^{-,+} G^{0,0*}_Y$.

Finally, we would like to show the covariant structure of the ghost
sector and extract the Feynman rules necessary for our calculation.
We introduce the mass eigenstates in $SU_L(2)\times U_Y(1)$
doublets:

\begin{equation}
C_Y=\left(\begin{array}{ccc} C^{0}_Y \\
C^-_Y
\end{array}\right) \ \ \ \ \ \bar{C}_Y=\left(\begin{array}{ccc} \bar{C}^{0}_Y \\
\bar{C}^-_Y
\end{array}\right).
\end{equation}
where the mass eigenstates are defined as
$C^{0}_Y=\frac{1}{\sqrt{2}}(C^4-iC^5)$ and
$C^{-}_Y=\frac{1}{\sqrt{2}}(C^6-iC^7)$, and similar expressions for
the antighost field $\bar{C}_Y$. The ${\cal L}_{FP}$ Lagrangian can
thus be written as follows:
\begin{eqnarray}
{\cal L}_{FP}&=&(D_\mu C_Y)^\dag (D^\mu
\bar{C}_Y)+\frac{g^2}{4}\Big[(Y^\dag _\mu \sigma^i
Y^\mu)(C^\dag_Y\sigma^i \bar{C}_Y)+3(Y^\dag_\mu
Y^\mu)(C^\dag_Y\bar{C}_Y)-4(Y^\dag_\mu C_Y)(Y^{\mu
\dag}\bar{C}_Y)\Big]\nonumber \\
&&+\frac{ig}{\sqrt{2}}Y^\dag_\mu M_CD^\mu
\bar{C}_Y+\frac{ig}{2}Y^\dag_\mu {\cal M}_C\bar{C}_Y-\frac{\xi
g}{2}\Big[\chi^{'0*}\chi^{'0}C^\dag_Y\bar{C}_Y+\chi^{'0}\Phi^\dag_YM_C\bar{C}_Y
-(C^\dag_Y\Phi_Y)(\Phi^\dag_Y\bar{C}_Y)\Big]\nonumber
\\
&&+\frac{i\sqrt{2}}{\xi}\Big[(\bar{M}_CC_Y+M_C\bar{C}_Y)^\dag (D_\mu
Y^\mu)-(D_\mu
Y^\mu)(\bar{M}_CC_Y+M_C\bar{C}_Y)\Big]\nonumber \\
&&-g\Big[\Phi^\dag_Y(\bar{M}_CC_Y+M_C\bar{C}_Y)\chi^{'0}+\chi^{'0*}(\bar{M}_CC_Y+M_C\bar{C}_Y)^\dag
\Phi_Y\Big]+{\rm H.c.}\nonumber \\
&&-\frac{1}{2}f^{\bar{a}bc}f^{cde}\bar{C}^{\bar{a}}\bar{C}^bC^dC^e,
\end{eqnarray}
where

\begin{equation}
M_C=\left(\begin{array}{ccc} \frac{1}{\sqrt{2}}(C^3+\sqrt{3}C^8) & \frac{1}{\sqrt{2}}(C^1-iC^2) \\
 \frac{1}{\sqrt{2}}(C^1+iC^2)&-\frac{1}{\sqrt{2}}(C^3-\sqrt{3}C^8)
\end{array}\right),
\end{equation}

\begin{equation}
{\cal M}_C=\left(\begin{array}{ccc} ({\cal D}^{3i}_\mu
+\sqrt{3}{\cal D}^{8i}_\mu)C^i &
({\cal D}^{1i}_\mu-i{\cal D}^{21}_\mu)C^i \\
 ({\cal D}^{1i}_\mu +i{\cal D}^{2i}_\mu)C^i&-({\cal
 D}^{3i}_\mu-\sqrt{3}{\cal D}^{8i}_\mu)C^i
\end{array}\right),
\end{equation}
where $i=1,2,3,8$, ${\cal
D}^{ij}_\mu=\delta^{ij}\partial_\mu-gf^{ija}A^a_\mu $ is the
covariant derivative given in the adjoint representation of
$SU_L(3)$, and $\bar{M}_C$ is obtained from $M_C$ after replacing
the ghost fields by the antighost fields. Under a $SU_L(2)\times
U_Y(1)$ unitary transformation $U$, $M_C$ transforms as $M_C \to
UM_CU^\dag$. The same is true for $\bar{M}_C$ and ${\cal M}_C$. It
is thus clear that ${\cal L}_{FP}$ is $SU_L(2)\times U_Y(1)$
invariant.

From  ${\cal L}_{FP}$, it is straightforward to show that the
Feynman rules for the trilinear and quartic couplings involving the
ghost fields and the SM gauge bosons have the same Lorentz structure
than those involving the pseudo-Goldstone bosons and the SM gauge
bosons, which stems from the fact that each sector is $SU_L(2)\times
U_Y(1)$ invariant by itself. The Feynman rules for the ghost
(antighost) fields are summarized in Fig. \ref{SFR} together with
the Feynman rules for the pseudo-Goldstone bosons. We see that the
trilinear vertices $WS^\dag S$ and $VS^\dag S$, with $S$ standing
for a commutative (pseudo-Goldstone boson) or anticommutative
(ghost) charged scalar, satisfy simple Ward identities:
\begin{equation}
k^\alpha \Gamma^{VS^\dag S}_\alpha=\Pi^{S^\dag
S^\dag}(k_2)-\Pi^{SS}(k_1),
\end{equation}
where $\Gamma^{VS^\dag S}_\alpha=(k_1-k_2)_\alpha$  and
$\Pi^{SS}(k_i)$ stands for the two-point vertex functions
$\Pi(k)=k^2-\xi m^2_Y$.

\section{Analytical results and discussion}
\subsection{A gauge invariant and gauge independent Green function
for the $WWV^*$ vertex} \label{cal}

The most general transverse CP-even vertex function for the
$W^+_\alpha(p-q)W^-_\beta(-p-q)V^*_\mu(q)$ coupling has the form
\cite{Bardeen:1972vi,Gaemers1979}
\begin{eqnarray}
\label{G_WWV} \Gamma^V_{\alpha \beta \mu}&=&-ig_V\Big\{A[2p_\mu
g_{\alpha \beta}+4(q_\beta g_{\alpha \mu}-q_\alpha g_{\beta
\mu})]+2\Delta \kappa_V (q_\beta g_{\alpha \mu}-q_\alpha g_{\beta
\mu}) +\frac{4\Delta Q_V}{m^2_W}\Big(p_\mu q_\alpha
q_\beta-\frac{1}{2}q^2p_\mu g_{\alpha \beta}\Big)\Big\}.
\end{eqnarray}
The longitudinal terms were ignored as they become negligibly small
when $V^*$ couples to a light fermion current, as in $e^-e^+\to WW$
scattering. The SM tree-level values are $A=1$ and $\Delta \kappa
=\Delta Q=0$.  While $\Delta \kappa$, and $\Delta Q$ are ultraviolet
finite at the one-loop level, $A$ is divergent and requires
renormalization.  The coefficients $\Delta \kappa_\gamma$ and
$\Delta Q_\gamma$, obtained from the  $WW\gamma$ vertex function,
determine the static electromagnetic properties of the $W$ boson,
namely, its magnetic dipole moment $\mu_W$ and its electric
quadrupole moment $Q_W$.

From the Feynman rules presented in Figs. 1 and 2, we can construct
all the Feynman diagrams contributing to the $WWV$ vertex at the
one-loop level, which we show in Fig. 3, and extract the $\Delta
\kappa_V$ and $\Delta Q_V$ coefficients. Although there is a similar
set of Feynman diagrams for the pseudo-Goldstone bosons and the
ghost fields, the unphysical particles only contribute to Eq.
(\ref{G_WWV}) via the triangle diagram. Moreover, it turns out that
the ghost (antighost) contribution is minus twice that of the
pseudo-Goldstone bosons, which is due to the separate $SU_L(2)\times
U_Y(1)$ invariance of these sectors. We have evaluated the loop
amplitudes in the Feynman-'t Hooft gauge via the Passarino-Veltman
reduction method \cite{Passarino:1978jh}. We have verified that the
bilepton contribution can be written exactly as Eq. (\ref{G_WWV}),
which means the $WWV^*$ Green function satisfies the Ward identity
$\Gamma^V_{\alpha \beta \mu} q^\mu=0$. It turns out that the form
factors are the same regardless of $V$, which means that, just as
occurs at the tree-level, $\Gamma^\gamma_{\alpha \beta \mu}$ and
$\Gamma^Z_{\alpha \beta \mu}$ only differ by the $g_V$ coefficient.
Thus $SU_L(2)\times U_Y(1)$ invariance is preserved at one-loop
level. The $g_V$ coefficient is given by
$g_V=e_V(Q^V_{Y^{0}}-Q^V_{Y^-})$, which, after inserting the charge
values, gives $g_V=e$ for $V=\gamma$ and $g_V=c_Wg$ for $V=Z$.

We now introduce the following short-hand notation $Q=2q$, $\hat
Q^2=Q^2/m^2_W$, $x_Y=m^2_Y/m_W^2$, $\beta(\hat Q^2)=6a/(4-\hat
Q^2)^3$, $a=g^2/96\pi^2$, and express the $\Delta \kappa_V$ and
$\Delta Q_V$ coefficients in terms of two- and three-point
Passarino-Veltman scalar functions:

\begin{eqnarray}
\label{Q_V} \Delta Q_V&=&2\beta(\hat Q^2)\Bigg[ \left(6x_Y
G(Q^2)+1\right)\hat Q^4- 2 \left(2\left( 2x_Y+1 \right)F(m_W^2,Q^2)
+
2x_Y F(0,Q^2) + 6\left(3x_Y-1 \right)G(Q^2) +3\right)\hat Q^2\nonumber\\
&+& 4\left( \left(8x_Y-5 \right)F(m_W^2,Q^2) +
8x_YF(0,Q^2) + 6\left(3x_Y-1 \right)G(Q^2)+5 \right)\nonumber\\
&+&8\left( 3F(m_W^2,Q^2) - 8x_YF(0,Q^2) +3\left( 1 - 4x_Y \right)
G(Q^2) -6 \right)\frac{1}{\hat Q^2}\Bigg]
\end{eqnarray}

\begin{eqnarray}
\label{k_V} \Delta \kappa_V&=&\beta(\hat Q^2)\Big[\left(F(m_W^2,Q^2)
- 4\,x_YF(m_W^2,0) - 6\,\left( 1 + 3\,x_Y \right)\,G(Q^2) -3
\right)\hat Q^4
+ 2\left( 2 + 16 x_Y F(m_W^2,0)\right.\nonumber\\
&+&\left.\left(13+16 x_Y\right)F(m_W^2,Q^2) -
3\left(1-3x_Y\right)G(Q^2) \right)\hat Q^2+ 32\left(1 - 2\,\left(
F(m_W^2,0)+2\,F(m_W^2,Q^2) \right)\right) \,x_Y \Big],
\end{eqnarray}
where $F(m^2,Q^2)=B_0(m^2,m^2_Y,m^2_Y)-B_0(Q^2,m^2_Y,m^2_Y)$ and
$G(Q^2)=m_W^2\,C_0(Q^2,m^2_W,m^2_W,m^2_Y,m^2_Y,m^2_Y)$, with the
scalar two-point $B_0$ and three-point $C_0$ functions given in the
usual notation. From here, it is clear that ultraviolet divergences
cancel up in the $W$ form factors.

Before the numerical evaluation, it is interesting to analyze the
static electromagnetic form factors, i.e. the scenario when $Q^2=0$.
Note that $\hat Q^2$ and $F(0,Q^2)$ vanish when $Q^2=0$, whereas
$G(0)=(2-F(m_W^2,0))/(1-4 x_Y)$ \cite{Stuart:1987tt}. After taking
the $Q^2\to 0$ limit in the last two terms of Eq. (\ref{Q_V}),
$\Delta Q_\gamma$ can be written as

\begin{eqnarray}
\Delta Q_\gamma &=&\frac{3 m_W^2 a}{2} \left(
\frac{1}{6 m_W^2(1-4x_Y)}\left(-21+48x_Y+3\left(1+\left(10-32x_Y\right)x_YF(m_W^2,0)\right)\right)\right.\nonumber\\
&+&\left.\left(8x_Y-3\right)\frac{\partial
B_0(Q^2,m_Y^2,m_Y^2)}{\partial Q^2}\Bigg|_{Q^2
=0}+3m_W^2\left(1-4x_Y\right)\frac{\partial
C_0(Q^2,m^2_W,m^2_W,m^2_Y,m^2_Y,m^2_Y) }{\partial Q^2}\Bigg|_{Q^2
=0}\right),
\end{eqnarray}
where L'H\^opital rule has been used for the indeterminate limit.
Since any three- and two-point scalar function and their derivatives
can be written as a combination of two-point functions
\cite{Stuart:1987tt}, we can write
\begin{equation}
\frac{\partial B_0(Q^2,m_Y^2,m_Y^2)}{\partial Q^2}\big|_{Q^2
=0}=\frac{1}{6m_Y^2},
\end{equation}
and
\begin{equation}
\frac{\partial C_0(Q^2,m^2_W,m^2_W,m^2_Y,m^2_Y,m^2_Y) }{\partial
Q^2}\big|_{Q^2 =0}=\frac{1}{6m_W^2m_Y^2}\frac{1-\left(5-6x_Y
\right)x_Y F(m_W^2,0)-5x_Y}{(1-4x_Y)^2}.
\end{equation}
We thus obtain
\begin{equation}
\Delta
Q_\gamma=\frac{4a}{1-4x_Y}\left(-1+\left(1-3\left(2x_Y-1\right)
\left(B_0(m^2_W,m^2_Y,m^2_Y)-B_0(0,m^2_Y,m^2_Y)\right)\right)x_Y\right),
\end{equation}
whereas $\Delta\kappa_\gamma$ can be obtained immediately from
(\ref{k_V})
\begin{equation}
\Delta \kappa_\gamma=\frac{3a}{2}\left(1 - 6x_Y
\left(B_0(m^2_W,m^2_Y,m^2_Y)-B_0(0,m^2_Y,m^2_Y)\right)\right).
\end{equation}
In the limit of very large $x_Y$ ($m_Y\gg m_W$) we obtain
\begin{equation}
B_0(m^2_W,m^2_Y,m^2_Y)-B_0(0,m^2_Y,m^2_Y)\simeq \frac{1}{6x_Y},
\end{equation}
which can be used to show that both $\Delta Q_\gamma$ and $\Delta
\kappa_\gamma$ vanish in the large bilepton mass limit, i.e.  these
coefficients are of decoupling nature. Furthermore, it is evident
that $\Delta Q_\gamma$ decreases more rapidly than $\Delta
\kappa_\gamma$ as $m_Y$ increases.

\begin{figure}
\centering
\includegraphics[width=4in]{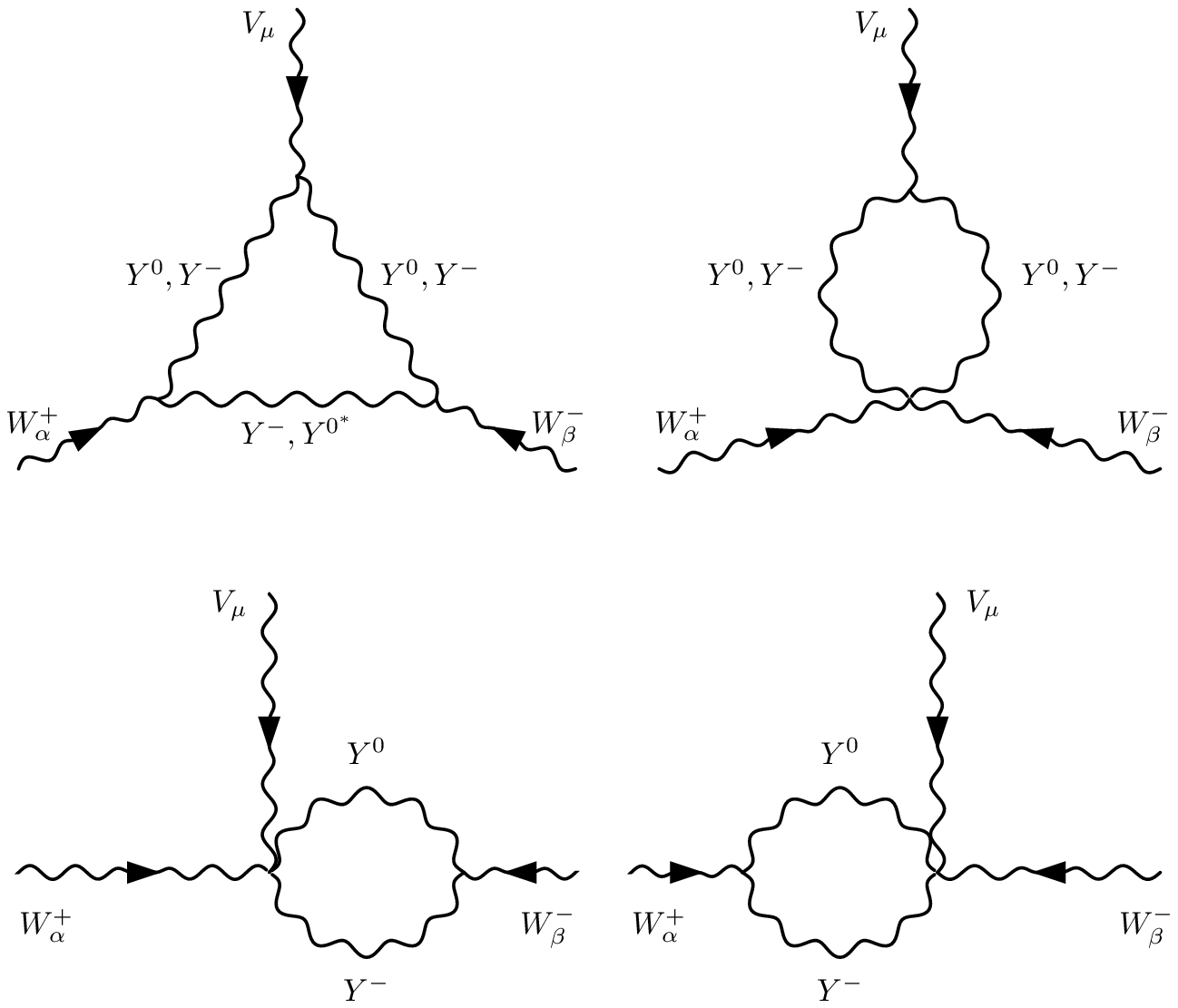}
\caption{\label{FIG4} Feynman diagrams for the $WWV$ vertex in the
$SU_L(2)\times U_Y(1)$-covariant gauge. The pseudo-Goldstone bosons
and ghosts contribute through an identical set of diagrams, but only
the triangle ones give a nonvanishing contribution to  $\Delta
\kappa_V$ and $\Delta Q_V$.}
\end{figure}

We emphasize that the above expressions account for the bilepton
effects on the $WWV^*$ vertex at the $w$ scale, when the bilepton
masses are degenerate and equal to $m_Y$. These results are thus
approximate but can give a realistic estimate of this class of
effects. We now proceed to the numerical evaluation and analysis.

\subsection{Numerical evaluation and discussion}

We now would like to analyze the behavior of  $\Delta Q_V$ and
$\Delta \kappa_V$ as functions of $Q^2$  and the bilepton mass,
which are the only free parameters which they depend on.

As far as $Q^2$ is concerned, we will focus on the values that can
be at the reach of the future planned linear colliders as this
vertex has the chance of being studied through $e^-e^+\to W^-W^+$
scattering. We thus consider the range 200 GeV $\le \sqrt{Q^2} \le$
2 TeV \cite{Accomando:2004sz}.

As for the bilepton mass, it is convenient to give a short account
on the existing bounds from both the theoretical and experimental
sides. First of all, we already mentioned that in the 331 model with
right-handed neutrinos the matching of the gauge couplings constants
at the ${SU_L(3)}\times {U_X(1)}$ breaking scale requires that
$\sin^2 \theta_W \le 3/4$ \cite{Foot:1994ym}, from which an upper
bound on the bilepton mass can be derived  after considering
radiative corrections to $\sin \theta_W$  and the most recent
experimental data \cite{Yao:2006px}. The respective bounds is
somewhat weak and contrasts with the case of the minimal 331 model,
which requires that $m_Y \lesssim 1$ TeV
\cite{Frampton:1992wt,Ng:1992st,Dias:2004dc} for consistency with
the theoretical bound $\sin^2 \theta_W \le 1/4$. Therefore, whereas
the bileptons predicted by the minimal 331 model could be searched
for via collider experiments in a near future, thereby confirming or
ruling out the model, the 331 model with right-handed neutrinos
would still leave open the door. We have also mentioned that because
of the symmetry breaking hierarchy, the bilepton mass splitting is
bounded by $\left|m_{Y\pm}^2-m_{Y^0}^2\right|\le m_W^2$. This means
that $m_{Y^0}$ and $m_{Y^\pm}$ cannot be arbitrarily different. In
fact, the heavier the bilepton masses, the closer its degeneracy.
Our approximation of mass degeneracy has thus much sense. We now
would like to examine the lower bounds on the bilepton masses. In
Ref. \cite{Ky:2000dj} it was argued that the data from neutrino
neutral current elastic scattering give a lower bound on the mass of
the new neutral gauge boson $m_{Z'}$ in the range of 300 GeV, which
along with the symmetry-breaking hierarchy yield $m_{Y^\pm} \sim
m_{Y_0} \sim 0.72\, m_{Z'}\ge 220$ GeV. A similar bound was obtained
in Ref. \cite{Long:1999yv} from the observed limit on the wrong muon
decay $R=\Gamma(\mu^-\to e^- \nu_{e} \bar{\nu}_\mu)/\Gamma(\mu^-\to
e^- \bar{\nu}_{e} \nu_\mu)\le 1.2\%$, which leads to $m_{Y^\pm}\ge
230\pm 17$ GeV at 90\% C.L. These lower bounds on $m_Y^{\pm}$ are in
agreement with that obtained from the latest BNL measurement on the
muon anomaly \cite{Ky:2000dj}. It is then reasonable  to consider
the range 200 GeV $\le m_{Y^0}\le$ 2000 GeV for our numerical
analysis. This will allow us to assess the behavior of the form
factors and get an estimate of their order of magnitude.   A word of
caution is in order here: it has been pointed out that some of the
existing bounds on the bilepton masses are too model dependent and
thus can be relaxed by considering extra assumptions
\cite{Pleitez:1999ix}, so a relatively light bilepton cannot be
ruled out yet.

\begin{figure}[!htb]
\centering
\includegraphics[width=5in]{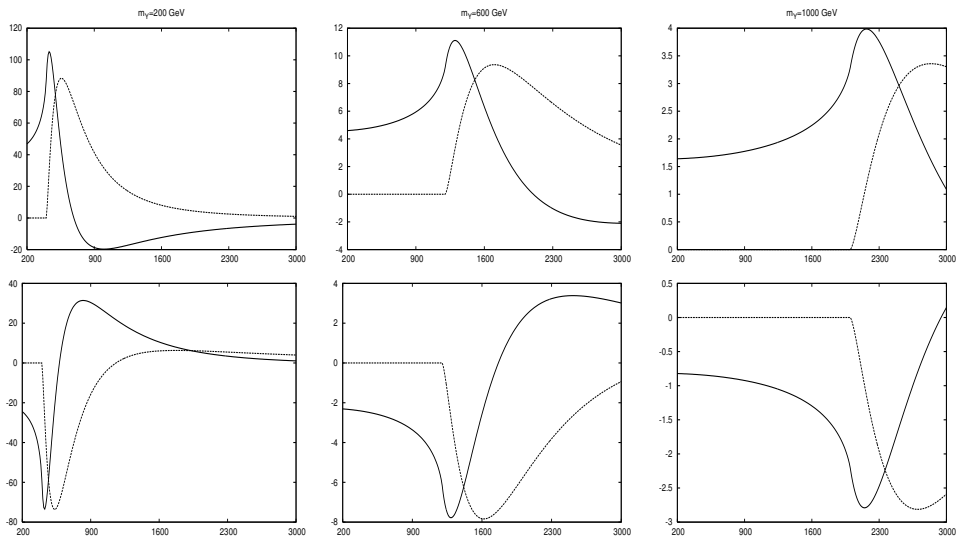}
\caption{\label{FIGFFQD} The $W$ form factors $\Delta Q_V$ (upper
plots) and $\Delta \kappa_V$ (lower plots) vs the center-of-mass
energy $E_{CM}=\sqrt{Q^2}$ for various values of the bilepton mass.
The form factors are in units of $10^{-6}$ and $E_{CM}$ is in units
of GeV. The solid (dashed) lines represent the real (imaginary) part
of the form factors.}
\end{figure}

\begin{figure}[!htb]
\centering
\includegraphics[width=5in]{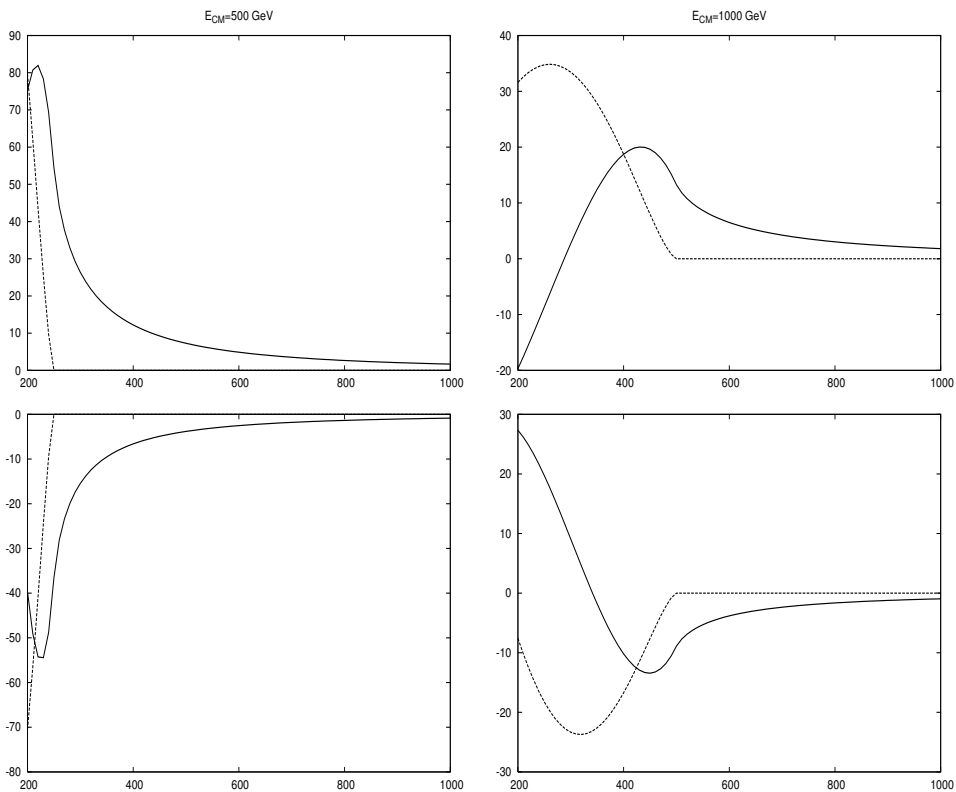}
\caption{\label{FIGFFMD} The $W$ form factors $\Delta Q_V$ (upper
plots) and $\Delta \kappa_V$ (lower plots) vs the bilepton mass
$m_Y$ for $E_{CM}=500$, and $1000$ GeV. The form factors are in
units of $10^{-6}$ and $m_Y$ is in units of GeV. The solid (dashed)
lines represent the real (imaginary) part of the form factors.}
\end{figure}

In Fig. \ref{FIGFFQD} we show the $W$ form factors $\Delta \kappa_V$
and $\Delta Q_V$ vs the center-of-mass energy $E_{CM}=\sqrt{Q^2}$,
i.e. the momentum carried by the virtual $V$ boson, for the bilepton
mass values $m_Y=200$, $600$, and $1000$ GeV. On the other hand, the
dependence of the $W$ form factors on the bilepton mass is shown in
Fig. \ref{FIGFFMD}, where we have plotted $\Delta \kappa_V$ and
$\Delta Q_V$ vs the bilepton mass for $E_{CM}=500$ and $1000$ GeV.

As far as the energy dependence of the form factors, from Fig.
\ref{FIGFFQD} it is clear that both $\Delta Q_V$ and $\Delta
\kappa_V$ are of the order of about $10^{-4}$ for a relatively light
bilepton with a mass of 200 GeV and a center-of-mass energy around
the energy threshold $E_{CM}=2m_Y$, where the form factors reach
their extremum values and develop an imaginary part. Below this
threshold, the form factors are purely real, which reflects the fact
that the bileptons that couple to the $V$ boson are necessarily
virtual, whereas the direct production of a bilepton pair becomes
possible for energies $E_{CM}\ge 2m_Y$. Although the direct
production of a bilepton pair would be preferred over the search for
their virtual effects, the latter would be useful for a high
precision test of the $WWV$ vertex. We can also see that the form
factors decrease by one order of magnitude when $m_Y= 500$ GeV and
by two orders of magnitude when $m_Y=1000$ GeV, which is depicted in
Fig. \ref{FIGFFMD}. For very high energies, the form factors have a
good behavior as they approach zero asymptotically after reaching an
extremum value above $E_{CM}=2m_Y$. Therefore, unitarity is
respected, which stems from the fact that the $WWV$ Green function
that we have obtained is gauge invariant. In Fig. \ref{FIGFFQD} we
can also observe that, except for the reversed sign, the curves for
$\Delta Q_V$ and $\Delta \kappa_V$ are very similar. For a
relatively light bilepton both form factors are of about the same
order of magnitude, but the magnitude of $\Delta Q_V$ decreases more
quickly as $m_Y$ increases.

It is interesting to compare our results with those obtained in the
SM and some of its extensions. As far as the SM is concerned, the
gauge boson contribution to the $\Delta Q_V$ form factor is of the
order of $10^{-3}-10^{-4}$ for $E_{CM}$ in the range $200-1000$ GeV,
whereas $\Delta Q_V$ is about one order of magnitude below
\cite{Argyres:1992vv,Papavassiliou:1993ex}. These contributions are
of the same order of magnitude than those of the fermion and scalar
sectors of the theory \cite{Argyres:1992vv}. As far as
supersymmetric models are concerned, the total contribution from the
unconstrained MSSM is about the same order of magnitude or larger
than the SM total contribution for some set of the values of the
parameters of the model \cite{Arhrib:1995dm}. Our results in the 331
model with right-handed neutrinos and relatively light bileptons,
with a mass around 200 GeV, is of the same order of magnitude than
the SM contribution. These results are similar to those obtained in
the minimal 331 model \cite{Montano:2005gs} and other weakly coupled
theories. Unless a very high precision is achieved in future
particle colliders, it would be very hard to discriminate the
contributions form different models to the $WWV$ vertex. However,
testing this vertex would still be useful for probing some
particular model once it has been confirmed by experimental data.

\section{Final remarks}
We have used a nonconventional quantization method inspired on the
BFM and BRST symmetry to analyze the effects of the new gauge bosons
(bileptons) predicted by the 331 model with right-handed neutrinos
on the of-shell $WWV^*$ vertex.  Hopefully, this class of effects
might be searched for at a future linear collider through $e^-e^+\to
V\to W^-W^+$ scattering. In particular, it has been pointed out that
CLIC would reach a sensitivity of about $10^{-4}$ in the measurement
of the so called form factors, $\Delta Q_V$ and $\Delta \kappa_V$,
characterizing the $WWV^*$ vertex \cite{Accomando:2004sz}. Following
the philosophy of the BFM, our method considers the bileptons as
quantum fields and the SM gauge bosons as classical fields.  A
nonlinear $SU_L(2)\times U_Y(1)$ invariant gauge-fixing is then
introduced for the bileptons, which are integrated out in the
generating functional. The result is a quantum action defined by a
$SU_L(2)\times U_Y(1)$ invariant Lagrangian\footnote{Actually, this
scheme is quite analogous to the effective Lagrangian approach, in
which an $SU_L(2)\times U_Y(1)$ effective Lagrangian is constructed
out of the SM fields to analyze the virtual effects of the heavy
fields, which have been integrated out.} out of which a gauge
invariant and gauge independent Green function can be obtained.  To
this end, we made use of the link between the diagrammatic method
known as the PT and the BFM, which establishes that the gauge
invariant and gauge independent Green function obtained via the PT
coincides with that obtained via the BFM as long as the Feynman-'t
Hooft gauge Feynman rules are used when calculating the latter. We
emphasize that our method is only approximate as the quantum action
is only invariant under $SU_L(2)\times U_Y(1)$ rather than
$SU_L(3)\times U_N(1)$. This method is suited to analyze the
bilepton effects at the $SU_L(3)\times U_N(1)$ breaking scale, when
their masses are still degenerate. The advantages of our calculation
scheme are twofold: the introduced nonlinear gauge-fixing term for
the bileptons allows one to remove several unphysical vertices,
which in turn allows one to get rid of several Feynman diagrams; on
the other hand, preserving the electroweak invariance turn the
calculation into a simple task as each sector of the theory gives a
gauge invariant contribution satisfying simple Ward identities by
its own. Once our method was applied, we obtained the form factors
$\Delta Q_V$ and $\Delta \kappa_V$, which were analyzed for several
values of the bilepton mass and convenient values of the momentum
carried by the virtual $V$ boson. It was found that the bilepton
effects on the $WWV^*$ vertex are of the same order of magnitude
than the SM and the minimal 331 model contributions, provided that
the bilepton mass is of the order of a few hundred of GeVs. For very
heavy bileptons, the respective contribution to the $WWV^*$ is
negligibly small. This indicates that it will be very hard to
discriminate between different class of effects on the $WWV^*$
vertex arising from distinct models. However, if a high precision is
achieved in future linear colliders, the $WWV^*$ vertex might serve
as a probe to some particular model by then confirmed by other
experimental data. The good behavior of the form factors at high
energies was also discussed, for it is a reflect of the gauge
invariance and gauge independence of the $WWV^*$ Green function
obtained via our quantization method.

\acknowledgments{We acknowledge support from SNI and Conacyt
(M\'exico).}

\end{document}